\documentclass[aps,prl,twocolumn,epsf,epsfig,amsmath]{revtex4}

\usepackage{latexsym}
\usepackage{hyperref}
\usepackage{times}
\usepackage{graphicx}
\usepackage{amsmath}

\usepackage{dcolumn}
\usepackage{bm}
\usepackage{ textcomp }
\usepackage{color}
\usepackage{amssymb}
\usepackage{xcolor}
\usepackage{easyReview}
\usepackage{mathdots}
\usepackage{float}
\usepackage{lineno}
\usepackage{todonotes}
\usepackage{array}
\newcommand{\beq}{\begin{eqnarray}}
\newcommand{\eeq}{\end{eqnarray}}

\begin{document}
\title{Fifty years of Wilsonian renormalization and counting}
\author{Philip W. Phillips}
\email[]{dimer@illinois.edu}
\affiliation{Department of Physics and Institute of Condensed Matter Theory, University of Illinois at Urbana-Champaign, Urbana, IL 61801, USA}

\begin{abstract}
Renormalization began as a tool to eliminate divergences in quantum electrodynamics but it is now the basis of our understanding of physics at different energy scales. I review its evolution with an eye towards physics beyond the Wilsonian paradigm.
\end{abstract}
\date{October 2022}


\maketitle

  Despite their microscopic differences, all simple fluids undergo a transition to the gas phase with identical universal characteristics. By systematizing the underpinnings of this universality,  Kenneth Wilson formulated a far-reaching renormalization group principle \cite{wilsonkondo} and in so doing established the tools for the modern understanding of phase transitions, critical phenomena, and quantum field theory.   

\section{Pre-Wilson field theory}

Before Wilson\cite{wilsonkondo} tackled the question of universality, quantum field theory had been developed through efforts to combine quantum physics with special relativity. However, this introduced the problem of vacuum polarization. 

 Since the 1930s\cite{serber} it was known that the interaction of electromagnetic fields with the continuous distribution of `negative energy' states (positrons) amends Coulomb's law with a logarithmic divergence to linear order in the fine structure constant.  The divergence obtains at short distances $r\ll\hbar/mc=3.86\times 10^{-13}m$, where $m_e$ is the electron mass and $c$ the speed of light.   

 Fortunately, this divergence can be eliminated by defining a new effective charge which will depend on the energy scale. It is from this dependence that the idea of  a `running' coupling constant emerges.    Murray Gell-Mann and Francis Low\cite{GML} showed that to all orders in the fine-strucure constant, the vacuum polarization at energy scales $\mu$ that are large relative to the mass of an electron modifies the coupling constant $g(\mu)$ in accord with the scale-invariant form
\beq
\psi\left(g(\mu)\right)=\psi(g(\mu'))\left(\frac{\mu}{\mu'}\right)^a,
\eeq
where $a$ is a number and $\psi$ is some function that are not important for this discussion. This result shows that as the energy scale is varied, the new coupling constant is related to the original one by a scale-invariant or self-similar scale factor, $(\mu/\mu')^a$.   Considering $\mu$ and $\mu'$ as infinitesimally separated leads to a differential equation that in its modern form is written:
\beq
\frac{dg}{d\ln\mu}= \beta(g).
\label{eq:beta-function}
\eeq

Years before this equation was derived, Heisenberg noted that the fine-structure constant $\alpha\approx1/137\approx 2^{-4}3^{-3}\pi$ to an accuracy of $10^{-4}$.  The essence of Eq.~\eqref{eq:beta-function} is that it is pointless to ruminate over any particular value for $\alpha$. Instead, because of the charge renormalization, the fine-structure constant depends on the energy scale at which it is measured, typically represented by the momentum transferred by the interaction.  

\begin{figure}
\begin{center}
\includegraphics[width=3.0in]{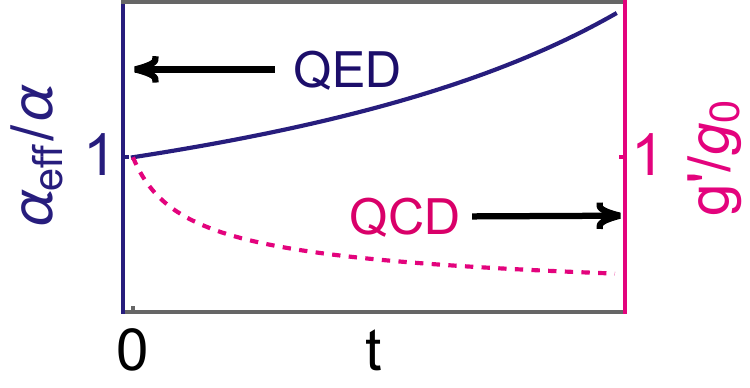}
\caption{\textbf{Running of the coupling constants}.  Illustrative plot of the $\beta$ functions for the coupling constants in quantum electrodynamics (QED), Eq. (\ref{alphaeff}) (QED) and quantum chromodynamics (QCD), Eq. (\ref{g'}) (QCD) as a function of the energy scale, $t$.   While both flow under renormalization, they do so in opposite directions.  QED becomes more strongly coupled at high energy while QCD does just the opposite.  At high energy, QCD is asymptotically free as the coupling constant vanishes. Confinement of the basic constituents, quarks and gluons, obtains at low energy in QCD.}
\label{qedqcd}
\end{center}
\end{figure}

In pure quantum electrodynamics (QED) consisting of a single photon field and an electron, the solution to Eq.~\eqref{eq:beta-function} predicts that the effective fine-structure constant,
\beq\label{alphaeff}
\alpha_{\rm eff}=\frac{\alpha}{1-\frac{\alpha}{3\pi}\ln\frac{-q^2}{e^{5/3} m_e^2}},
\eeq
depends explicitly on the transferred momentum, $q$, where $-q^2>0$ is an increasing function of energy. This behaviour was directly observed in the Large Electron-Positron (LEP) collider in 1994. While a full treatment with all the leptons and quarks is necessary to obtain the complete flow of $\alpha_{\rm eff}$ from  $\alpha_{\rm eff}(M_W^2)=1/128$ ($M_W$ the mass of the W-boson) to its low-energy value of approximately $1/137$, Eq.~\eqref{alphaeff} is sufficient to capture the deviation from the naive expectation that the local quasi-instantateous physics and hence only the bare parameters in the Lagrangian should matter in the high-energy limit.  This is not borne out in QED.  Fig. (\ref{qedqcd}) depicts that quantum chromodynamics (QCD) --- the theory of strong interactions between gluons and quarks --- stands in contradistinction to QED.  This is one of the great triumphs of Wilson's renormalization approach\cite{GW}.
 
While a theory with photon fields is naturally scale invariant, QED tells us that once matter is included, such scale invariance is lost by virtue of the running of the charge manifested in Eq.~\eqref{alphaeff}.   Nonetheless, the presence of a logarithm in the $\beta-$function reflects, to quote Wilson, `` a problem lacking a characteristic scale.''\cite{wilsonkondo} In fact, a similar logarithm arises in the theory for the ground-state energy of an electron gas, which features a Fermi sea of positive-energy electron states rather than the negative-energy positrons of the vacuum. 

How are these two features of QED compatible?  In QED and elementary particle theory in general, the only discernible energy scale is set by the rest mass of the constituents.  Integrals of the form 
\beq
\int_{m_ec^2}^\infty\frac{dE}{E}
\eeq
are logarithmically divergent precisely because all energy scales above $m_ec^2$ contribute equally.  Consider a scale $E'>mc^2$.  The contribution to the integral from $E'$ to $2E'$ is simply $\ln 2$, independent of the scale $E'$.  

In practice, all field theories are defined up to a high-energy cutoff or equivalently a short-distance scale.  Precisely the role played by the high-energy (short-distance) cutoff in an analysis of field theories lies at the heart of Wilson's approach to renormalization.   As we will show, what Wilson clarified is that low-energy theories depend on short-distance physics through operators classified as relevant, marginal and, in some cases, irrelevant depending on the energy scale being probed.  It is from this dependence that universality arises.

To set this up, we note that in field theory, it is not the value of a field at any point that matters but rather correlation functions of the underlying fields.  A key precursor to Wilson's work was the Callan-Symanzik\cite{CS1,CS2} equation,
\beq
\left[\mu\frac{\partial}{\partial\mu}+\beta(g)\frac{\partial}{\partial g}+n\gamma(g)\right]G^n(\mu,g,\gamma)=0,
\eeq
which established that any $n-$point correlation function $G$ is independent of the cutoff through two universal functions that communicate the shift in the coupling constant, $\beta(g)$, and the field strength, $\gamma(g)$, in such a way to counteract the shift made in the energy scale, $\mu$. 

\section{Block renormalization}
The story of renormalization prior to 1971 is more tied to removing infinities that arise in computing Feynman graphs than it is to some universal physical principle involving collective degrees of freedom.  Wilson provided\cite{wilsonkondo} this missing link by focusing on how systems with fluctuations on all length scales, such as a boiling pot of water, can be studied without forgoing locality.   

One of the simplest models featuring a phase transition is the Ising model for the onset of ferromagnetism. In this model, spins with either an up or down degree of freedom occupy sites with a separation of $a$ on a $d$-dimensional lattice and interact with nearest-neighbour interactions. 
In this context, Leo Kadanoff introduced a block coarse-graining renormalization scheme \cite{kadanoff} for the Ising model in which the entire system is divided into cells of edge length $ba$ ($b>1$). This approach provides an operational way to build in fluctuations smaller than the correlation length, $\xi$.   

A new coarse-grained spin variable is introduced to represent the average of the $b^d$ spins in each block.  The Hamiltonian can then be rewritten to take the same form at each iteration as long as the block spins are normalized by a rescaling factor to maintain the up-down or $Z_2$ Ising symmetry.  

The major conceptual leap in this approach is the assumption that the  blocks, like the underlying spins,  only have nearest neighbour interactions. An initial system of $N$ spins has $Nb^{-d}$ effective spins after blocking, each separated by $ba$. The correlation length $\xi$ can be represented either in units of the initial lengthscale $\xi=\xi_1\times a$ or the blocked lengthscale $\xi=\xi_b\times ba$. The rescaled correlation length $\xi_b$ is smaller than the correlation length at the initial scale $\xi_1$:
\beq
\xi=\xi_b\times(ba)=\xi_1\times a\Rightarrow \xi_b=\frac{\xi_1}{b}.
\eeq
Consequently, the corresponding rescaled Hamiltonian for the $b^{\rm th}$ iteration $H_b$ lies further away from a critical point --- where the correlation length diverges --- than the initial Hamiltonian $H_1$.  

This is reflected by the rescaled temperature and magnetic field parameters in the model, $t_b$ and $h_b$, respectively.  Let $t$ and $h$, be the bare values of the temperature and magnetic field, respectively.  A key assumption in the renormalization group procedure is that after rescaling, these quantities satisfy power-law scaling laws, $t_b=t b^{y_t}$ and $h_b=h b^{y_h}$ where $y_t$ and $y_h$ are both positive and can only be determined from the full renormalization transformations.   This leads to a series of recursion equations that ultimately make it possible to sample the infinite hierarchy of fluctuations with only a finite number of degrees of freedom at each step. 

\section{Power counting}

A revolution came with Wilson's\cite{wilsonkondo} momentum-space translation of the Kadanoff\cite{kadanoff} real-space coarse graining. It represented the degrees of freedom in the Ising model as fields in continuous space. This approach brought the physics of critical phenomena into quantum field theory, and through renormalization, established what field theory looks like in the statistical continuum limit.  

The notion of renormalizabilty is in general ill-posed as normally stated, as one must mention the space of operators within which a theory is renormalizable. 
More explicitly, consider a certain theory described by a classically local action $S(\phi_i)$ of some fields $\phi_1,\cdots ,\phi_n$. Suppose the field theory is valid up to some energy scale $E_0$ and we seek a theory valid for energies below this scale, $E<E_0$.  To do this, one introduces a cutoff scale $\Lambda<E_0$ and `integrates out' fields whose energy is higher than $\Lambda$ to obtain an effective action $S_\Lambda$ that depends only on low energy degrees of freedom. This is the energy- and momentum-space equivalent process to the blocking step of Kadanoff's procedure.

Operationally, this is done by splitting the field into high and low-energy components
\beq
\phi(\omega)=\left\{\begin{array}{ll}\phi_L(\omega), &\omega<\Lambda\nonumber\\
\phi_H(\omega), &\omega>\Lambda\end{array}\right.,\
\eeq
where $\omega$ is the energy, and performing an integration over the high-energy modes in the partition function to obtain the effective low-energy theory: 
\beq 
 \int D\phi e^{i S(\phi)}= \int D\phi _L e^{i S_\Lambda (\phi_L)}
 \eeq
 where 
 \beq
 S_\Lambda (\phi_L)= -i\log\left( \int  D\phi _H e^{i S(\phi_L , \phi_H)}\right)
 \eeq
 is the outcome of the integration.

 In the analysis of running coupling constants there are special values known as fixed points for which $\beta=0$ in Eq.~\eqref{eq:beta-function}. Defining $S_*$ as the action at a particular fixed point, one can write the action for a different set of parameters as 
 \beq \label{seff}
 S_\Lambda =S_*+ \int d^d x \sum _i g_i \mathcal O _i
 \eeq
 for some set of field operators $ \mathcal O _i$ that are local despite the integration of high frequency fields, because we focus on fields with $\omega <\Lambda$.  
 
 As with the block renormalization approach, we can consider the behaviour of the model under length rescaling,
\beq
x^\mu\rightarrow \lambda^{-1} x^\mu.
\eeq
If, under such a transformation, an operator $\mathcal O(x)$ can be written as
\beq
{\mathcal O}(x)=\lambda^{d_{\cal O}}{\mathcal O(\lambda^{-1}x}),
\eeq
we interpret $d_{\cal O}$ as the dimension of ${\mathcal O}$. 

Under a rescaling, the action can be organized based on the 
 exponent of $\lambda$ in each term, a procedure known as power counting.  In the $\lambda\rightarrow\infty$ limit, each operator will either remain invariant, vanish or blow up. The rule is as follows. Because of the $d$-dimensional spacetime measure in the action, operators with $d_{\mathcal O}-d>0$ are irrelevant and do not influence the low-energy physics. Relevant operators correspond to $d_{\mathcal O}-d<0$. 

Operators with $d_{\mathcal O}-d=0$ are marginal. In these cases,  all scales are important and such operators are the origin of logarithms in the $\beta$ function.  

The core of renormalization is in the observation that there is a dimension $D$ above which the operators are irrelevant. Furthermore, the number of local operators $\mathcal O_i$ whose dimension is less than (or equal) to $D$ is finite.  This obtains because classically local operators are polynomials in the fields $\phi$ and their derivatives.  Since there are finitely many of these, one can make sense of  such theories.  Wilsonian renormalization rests on the simple principle that  the low-energy physics is determined only by the relevant or marginal interactions or in rare cases, irrelevant couplings but only at low enough scales.  That the details of renormalization are determined by the dimension of operators rather than the nature of the microscopic features of the interactions or the cut-off is the origin of universality in the Wilsonian approach.

There are subtleties\cite{polchinski} in evaluating $S_\Lambda$, which typically has to be performed perturbatively. However, these can be overcome by a slight recasting\cite{polchinski} of the problem set forth by Wilson. 
We can imagine integrating out high-energy modes one small energy slice at a time. First we remove the modes with energies in the range $\Lambda>\omega>\Lambda-d\Lambda$, then $\Lambda-d\Lambda>\omega>\Lambda-2d\Lambda$ and so on. At each stage the effective action $S_\Lambda$ changes, which is described by the Wilson equation,
\beq
\frac{\partial S_\Lambda}{\partial\Lambda}={\cal F}(S_\Lambda),
\eeq
where ${\cal F}$ is a well-defined functional that can be calculated.  

As the Wilson equation represents a flow in an infinite dimensional space, examining its properties for a range of operators can be accomplished entirely from the eigenvalue spectrum.  Irrelevant operators correspond to negative eigenvalues, which represent benign converging flows.  If the functional is linearized around zero-coupling, the eigenvalues are precisely the numbers $d_{\cal O}-d$ obtained from power counting.  As ${\cal F}(S_\Lambda)$ is a smooth function of the couplings, there is no place\cite{polchinski} for singularities to obtain especially since we are performing a path integral over a narrow range of energy with both a low and high energy cutoff.  Hence, if an eigenvalue is negative in the free theory, the same holds for the interacting theory.  Power counting then rules even if the dimension can change at strong coupling, for example in the  Thirring model; hence the claim of marginality or relevance is the crux of the matter.  


\section{The $\beta$ function}

The evolution of the action as high momentum states are integrated out is precisely what is described by the running of the coupling constants in the $\beta$ function.  What Wilson added beyond the Gell-Mann/Low flow equation, Eq.~\eqref{eq:beta-function}, is that the $\beta$ function is governed by power-counting, coupled with integration of the high-energy modes and rescaling.

In the theory of quantum chromodynamics (QCD), perturbative treatment of non-Abelian Yang-Mills gauge theories\cite{GW,politzer} yields a $\beta$ function of the form 
\beq\label{g'}
\beta(g)=-b g^3\rightarrow g'^2=\frac{g_0^2}{1+2bg_0^2 t},
\eeq
where $t$ is proportional to the energy transferred and $b$ is a numerical constant.  At high energies, $t\rightarrow\infty$, the coupling constant flows to zero, producing the phenomenon known as asymptotic freedom whereby quarks and gluons become weakly interacting and treatable using perturbation theory. The opposite is true at low energies, where instead confinement of quarks and gluons takes places, producing a divergence of the coupling constant and the general breakdown of the whole perturbative scheme\cite{GW,politzer}.

Similar phenomena occur for the seemingly unrelated problem of a localized magnetic spin engaging in spin-flip scattering with a non-interacting band of conduction electrons, which is known as the Kondo problem. The spin-flip scattering operator is marginal, and the coupling strength flows from an initial value of $g_0$ according to the $\beta-$function
\beq
\beta(g)=g^2\rightarrow g'=\frac{g_0}{1-g_0\ln E_0/E},
\eeq
where $E_0$ and $E$ are the initial and final energy scales, respectively. 
If the initial interaction is antiferromagnetic, $g_0<0$, the flow is towards strong coupling,  yielding a divergent coupling constant signaling the formation of a bound state between the impurity and the conduction electrons.


In both QCD and the Kondo problem,  the formation of new degrees of freedom in at low energies obtains through a cross-over rather than a phase transition.  A key triumph of Wilson's  treatment of the Kondo problem is that it captured the universal scaling that transpires below a characteristic temperature $T_k$ at the $g\rightarrow-\infty$ fixed point, where the local moment is completely screened.  As fixed points are characterized by scale invariance, the Kondo temperature is obtained by imposing the scale-invariant condition $D\partial T_k(g)/\partial D=0$, where $D$ is the bandwidth of the conduction electrons.  The universal physics of all properties such as resistivity, magnetism, and thermodynamics below the cross-over scale $T_k$ is a consequence of this scale-invariant condition at the strongly coupled fixed point of the theory.  


\section{Beyond Wilson}

Fifty years on, it is natural to wonder if there is any physics beyond the Wilsonian paradigm. This could arise from a theory in which as the high energy is probed features emerge in the scattering matrix that are distinct from the poles that correspond to particles. For example, in the case of a doped Mott insulator\cite{dzy} scattering matrix zeros describing non-propagating or incoherent degrees of freedom have been identified.  

Before we get to the Mott problem, consider quantum gravity.  As  pointed out previously\cite{leigh}, probing high energy in a gravitational theory should produce black hole information which cannot be represented as simple poles in the scattering matrix. This would lead to a failure of the Wilsonian separation of energy scales and provide an example of ultraviolet/infrared (UV/IR) mixing, which is well studied in the context of non-commutative field theories\cite{witten}.  Precisely how such UV/IR mixing plays out in a theory of quantum gravity remains an open question.

To date our best understanding of quantum gravity stems from gauge-gravity duality, also known as the AdS/CFT conjecture\cite{maldacena}.  But even this conjecture has an effective Wilsonian interpretation at its core arising from the locality of energy in the $\beta$-function.
The central claim of the gauge-gravity duality is that some strongly coupled conformally invariant field theories in $d$-dimensions are dual to a theory of gravity in a $d+1$ spacetime that is asymptotically described by an anti de Sitter (AdS) metric parameterized by 
 \beq\label{adsmetric}
 ds^2=\frac{R^2}{z^2}\left(\eta_{\mu\nu}dx^\mu dx^\nu+dz^2\right),
 \eeq
 where $R$ and $z$ are the radius and radial coordinates of the AdS spacetime, respectively.
This spacetime is invariant under the transformation $x_\mu\rightarrow \Lambda x_\mu$ and $z\rightarrow z\Lambda$ and hence satisfies the requisite symmetry for the implementation of the gauge-gravity duality, although not the full symmetry of the conformal group.  The conformal field theory is viewed as lying on the $z=0$ boundary of the AdS spacetime.

\begin{figure}
\begin{center}
\includegraphics[width=3.0in]{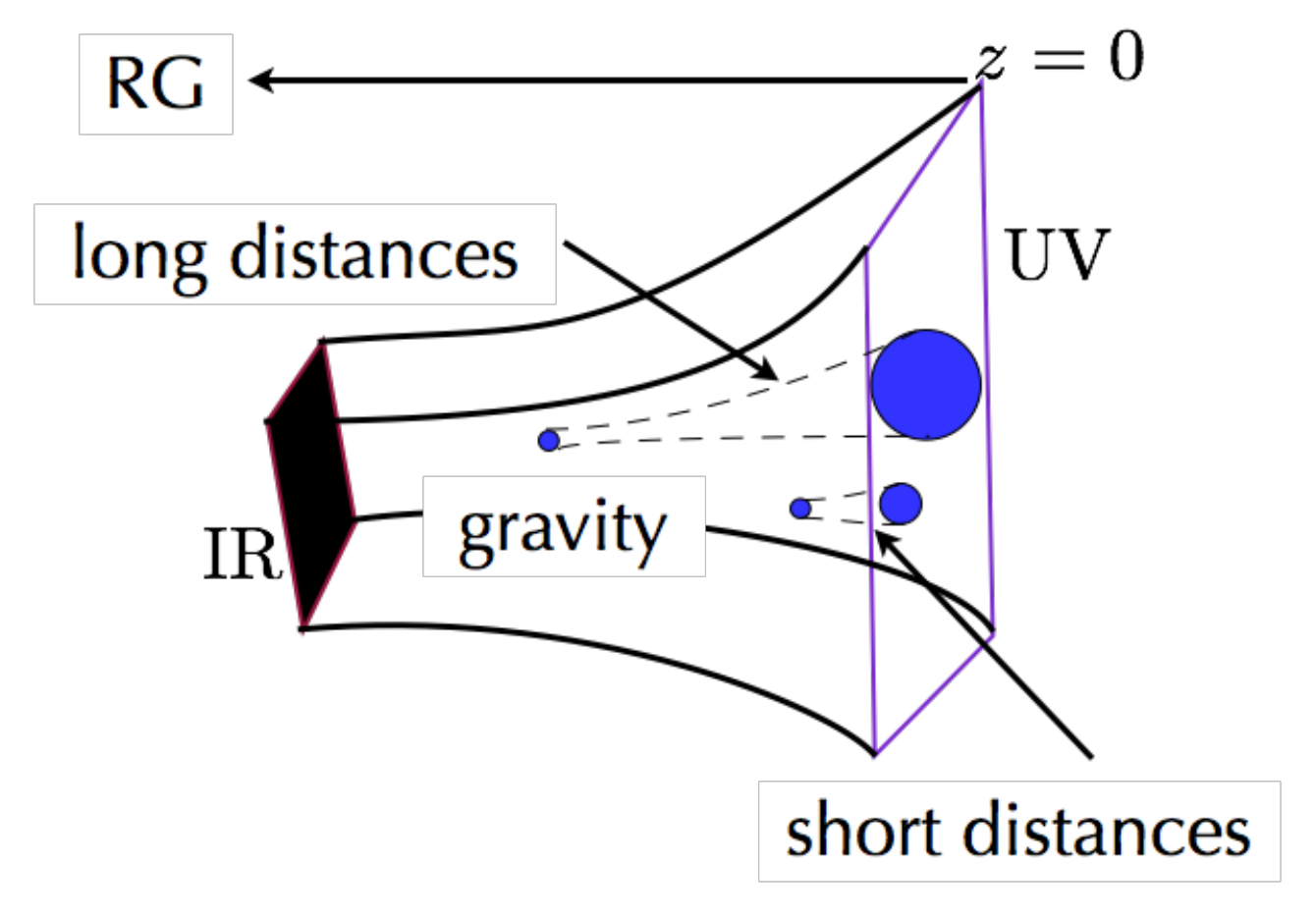}
\caption{  \textbf{Geometrical representation of the key claim of the gauge-gravity duality}.  A strongly coupled field theory lives at the boundary, at the high-energy UV scale.  The horizontal $z$-direction (the extra dimension in the gauge-gravity duality) represents the running of the renormalization scale.  This is illustrated by the two projections at different values of $z$ in the space-time.  Because the spacetime is asymptotically hyperbolic, larger values of $z$ lead to a larger projection of the boundary theory and a full running of the renormalization scale amounts to the construction of the low-energy, infrared (IR) limit of the original strongly coupled UV theory. }
\label{rgegr}
\end{center}
\end{figure}

Our current understanding of the radial coordinate $z$ is that it represents the flow in the energy scale during renormalization.   The scale change, $x_\mu\rightarrow \Lambda x_\mu$ increases the radial coordinate, $z\rightarrow z\Lambda$.   Consequently, moving towards greater $z$ in the bulk of the geometry increases the corresponding projection onto the boundary, as depicted in Fig. (\ref{rgegr}).  The limit of $z=\infty$ therefore represents the full low-energy or IR limit of the strongly coupled theory. The AdS/CFT conjecture thereby provides a complete geometrization of the renormalization group procedure.

The second area where a possible breakdown of the Wilsonian paradigm might arise is the strange metal\cite{science} phase in the cuprate superconductors which all start out as Mott insulators. In the strange metal, the resistivity increases way beyond the limit set by a scattering length determined by the physics of the underlying lattice constant.  As such a length scale determines the cutoff for particle scattering, the strange metal with its non-saturating resistivity requires physics beyond the Fermi liquid quasiparticle picture.  In terms of the standard field theory for a Fermi surface\cite{polchinski,shankar}, no operator in any effective Lagrangian exists that can account for a non-saturating resistivity from the lowest temperatures to temperature scales where the particle-picture breaks down.
This physics ultimately arises from the incoherent part of the spectrum, that is, zeros  of the scattering matrix, which has no Wilsonian formulation, at present.  So indeed, these two examples indicate that much physics possibly lies beyond the Wilsonian paradigm.

\textbf{Competing interests}
The author declares no competing interests.


\textbf{Acknowledgements} 

This work was funded through NSF DMR-2111379. I thank R. Leigh and Gabriele La Nave for always insightful conversations, Jinchao Zhao for help with Fig. 1 and Orlando Alvarez for his lucid renormalization group class at Berkeley in 1982.
 
\bibliography{rg}

\end{document}